\documentclass[prl,twocolumn]{revtex4-1}
\usepackage{amsmath,amsfonts,graphicx,hyperref}
\usepackage{wasysym}

\newcommand{\be}{\nopagebreak[3]\begin{equation}}
\newcommand{\ee}{\end{equation}}
\newcommand{\ba}{\nopagebreak[3]\begin{eqnarray}}
\newcommand{\ea}{\end{eqnarray}}
\newcommand{\fracs}[2]{{\textstyle\frac{#1}{#2}}} 

\begin{document}
 \title{Evidence for Maximal Acceleration and Singularity Resolution\\ in Covariant Loop Quantum Gravity}

\date{July 3, 2013}

\author{Carlo Rovelli}
\email{rovelli@cpt.univ-mrs.fr}
\affiliation{CPT, CNRS UMR7332, Aix-Marseille Universit\'e and Universit\'e de Toulon, F-13288 Marseille, EU}

 \author{Francesca Vidotto}
 \email{fvidotto@science.ru.nl} 
 \affiliation{Radboud University Nijmegen, \ Institute for Mathematics, Astrophysics and Particle Physics, 
Mailbox 79, P.O. Box 9010, 6500 GL Nijmegen, The Netherlands
\vskip1pt}
 

\begin{abstract}                
\vskip1pt
\noindent 
A simple argument indicates that covariant loop gravity (spinfoam theory) predicts a maximal acceleration, and hence 
forbids the development of curvature singularities. This supports the results obtained for cosmology and black holes
using canonical methods.  
\end{abstract}


\maketitle

\section{Introduction}

The singularities that appear in the solutions of the field equations of general relativity are symptoms of the limits of validity of the classical theory, which disregards quantum effects. Their study is a testing ground for quantum gravity.  In general, quantization yields discreteness, and quantization of gravity yields quanta of space with minimal size in the Planck regime \cite{Bronstein:1936kx,Bronstein:9vn}. In loop quantum gravity \cite{Rovelli:2004fk,ThiemannBook} this follows from the fact that geometrical quantities are described by operators that have discrete spectra \cite{Rovelli:1994ge,Ashtekar:1996eg}. But this result does not appear to be sufficient to remove the singularities since it is kinematical, while dynamics plays a role in the development of   singularities.  

Dynamics can be formulated in two ways in loop quantum gravity: in the canonical framework, via the definition of the quantum Hamiltonian constraint, and in the covariant framework, in terms of transition amplitudes expressed as ``sums over geometries" called spinfoams. 

Loop quantum cosmology \cite{Ashtekar:2011ly} has  studied the problem of the cosmological singularities extensively using canonical methods, providing strong elements of  evidence that these are eliminated by quantum effects.  The theory is based on a quantization of classical cosmological models, plus  ingredients imported from the full loop theory, assumed to capture the relevant quantum effects on spacetime geometry.  In particular, elements of the regularization of the Hamiltonian operator, such as the expression of curvature in terms of holonomies, or the inverse volume in terms of commutators, are taken to the reduced theory, and lead to the singularity resolution, also in rather generic cosmological contexts \cite{Singh:2009mz,Singh:2010qa}. Using the canonical loop techniques, indications have been obtained that black hole singularities as well are resolved {\cite{Modesto:2004xx,Ashtekar:2005cj,Ashtekar:2005qt,Gambini:2008bh,Gambini:2013qf}.

In recent years, the covariant version of the dynamics of loop gravity has developed extensively  \cite{Rovelli:2011eq}.  In this formalism manifest local Lorentz invariance can be maintained.  Can the results of loop cosmology be recovered from elements of the covariant theory? This would provide support to the reliability of the approximation that grounds loop cosmology.  In this Letter we provide some evidence that they can. This is our first result. 

We do not employ the full machinery of spinfoam cosmology \cite{Vidotto:2011qa,Bianchi:2010zs}. The key to our derivation relies on a core aspect of the covariant approach: the proportionality between generators of boosts and rotations  \cite{Rovelli:2011eq}.  This ties space-space and space-time components of the momentum conjugate to the gravitational connection and transfers the discretization of the area spectrum to a discretization of a suitable Lorentzian quantity, which, we show, is related to acceleration.  The mechanism indicates the existence of a maximal acceleration. This, in turn, yields a bound on the curvature and on the energy density in appropriate cosmological contexts, supporting the results in loop quantum cosmology and for black holes.

Maximal acceleration is a long expected quantum-gravitational phenomenon \cite{Caianiello:1981jq,Brandt:1988sh,
Toller:2003zu} and we regard the appearance of an indication of this phenomenon in loop gravity as our second and main result. Unlike other approaches, here maximal acceleration is compatible with local Lorentz symmetry, for the same reasons for which a minimal length is compatible \cite{Rovelli:2002vp,Rovelli:2010ed}.

The derivation also sheds  some light on the question of whether singularity resolution is kinematical or dynamical.  On the one hand, it is related to the discretization of the intrinsic geometry of space, which is a kinematical phenomenon, independent on the specific of the dynamics (like angular momentum quantization in quantum mechanics). On the other hand, it involves an analysis of the dynamics, to see how it reflects on the curvature or the energy density. Here we show that this ambiguity can be seen under a different light in the covariant theory. Singularity resolution is tied to the existence of a discrete spectrum, but it is the spectrum of a spacetime, not a spacial quantity.

\section{Acceleration}

Let us start from a worldline of constant acceleration $a$ in Minkowski space.  Any such worldline determines a spacetime point $H$ which is at equal four-distance $\ell=1/a$ from all the points of the worldline. $\ell$ is the distance of the horizon seen by the accelerated observer.  We work in units where the speed of light is unit. Pick a point $P$ and let $P'$ be the point at a hyperbolic angle $\eta$ from $P$.  Consider the region $\cal R$ bounded by the portion of the trajectory  from $P$ to $P'$ and the straight lines from $P$ to $H$ and from $H$ to $P'$; see Fig. 1.

 \begin{figure}[h]
\includegraphics[width=38mm]{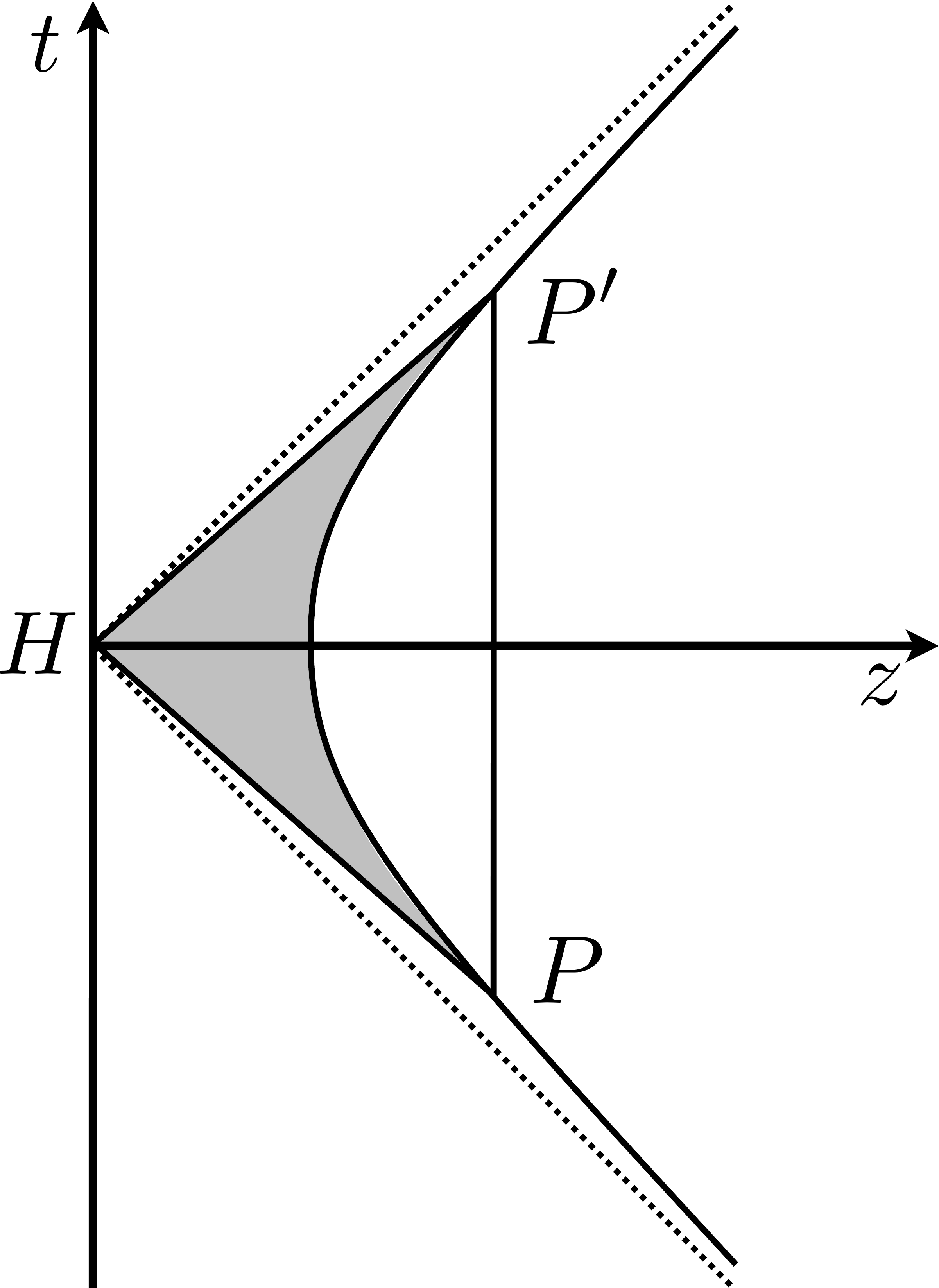}
			\caption{$\cal R $ is the shaded region.}
			\end{figure}
			
Consider the (Lorentzian) area of $\cal R$. This is easy to compute. Taking the origin of the coordinates in $H$ and the middle point between $P$ and $P'$ on the $z$ axis, the trajectory is given by the hyperbole 
\be
z(\eta)=   \ell \cosh(\eta) , \hspace{2em} t(\eta)=  \ell \sinh(\eta),  \label{time}
\ee
and the two points are located at the hyperbolic angles $\pm\frac\eta2$ respectively. The area of the shaded region $\cal R$ can be obtained from the difference  $A= A_{\triangle} - A_{\Leftcircle} $ between the area of the triangle $HPP'$ 
\be
A_{\triangle} = 2\ \frac12\, t(\fracs\eta2)z(\fracs\eta2)=\ell^2{\cosh(\fracs\eta2)\sinh(\fracs\eta2)}
\ee
and the area of the half-moon delimitated by the trajectory and the straight line from $P$ to $P'$
\be
A_{\Leftcircle} = { 2}  \int_{0}^{\fracs\eta2}\! \!   t(\tilde\eta) \, dz(\tilde\eta)=  \ell^2  \int_{0}^{ \fracs\eta2} \! \!\, \sinh^2(\tilde\eta)\  d\tilde\eta
\ee
which gives\\[-20pt]
\be
A=\frac\eta2\ell^2.
\ee
This is easy to understand geometrically. The area element $dz\wedge dt$ is Lorenz invariant. Therefore we can partition $\cal R$ into infinitesimal triangles with basis $ds=\ell d\eta$ and a vertex in $H$ and compute the area of each in its proper frame, which gives $dA=\frac12\ell^2 d\eta$. 

In particular, consider the ``equilateral" region  $\cal R$ where the proper length of the trajectory from $P$ to $P'$ has length $\ell$ (like the two other sides).   In this case $\eta=1$ and the area is  
\be
A=\frac12\ell^2=\frac{1}{2a^2}\,.\label{area}
\ee
Acceleration is a measure of the curvature of a timelike worldline. It is the Lorentzian analog of the curvature of a line in space, which, in turn, is determined by the size of the osculating circle, and in particular its area (or the area of the region wiped by the radius for an arc of the same length as the radius, as above).  Thus the area $A$ measures the acceleration $a$.

This measurement of the acceleration has a simple operational interpretation.  Say we are on the Earth's surface and we want to measure our acceleration with respect to an inertial frame (Galileo's measure).    An elegant way, in principle, is to throw a {clock} upward, and compare the time $s$ it takes to fall back, measured by a clock in our hands, with the time $t$ measured by the falling clock itself. A moment of reflection shows that this measure is precisely described by the math above, where the accelerated trajectory is ourselves and the freely falling trajectory of the falling clock (a geodesic) is the straight line from $P$ to $P'$.  Given the measured values $t$ and $s$ of the two clocks, we can get the acceleration $a$ from
\be
at= \sinh(a s).
\ee
Choosing a run where $t=s \sinh{1}$, amounts to taking $s=\ell=1/a$, and the area is as in \eqref{area}.   That is, the area of $\cal R$ is $\frac12$ the square of the reading of the clock in our hands, when the flying clock is slower by a factor $\sinh{1}$. 

\section{Gravity}
The action of general relativity can be expressed as
\be
S[e,\omega]= \int B[e] \wedge F [ \omega ]
\label{action}
\ee
where $F[\omega]$ is the curvature form of the spin connection $\omega$, $B$ is the two-form
\be
B[e]=
\Big(\!\ast +\frac1\gamma\,  \Big) \Big(   e \wedge e \,\Big)~,
\ee
$e$ is the tetrad and the star $\ast$ denotes the Hodge dual. For  notation and details, see Ref. \cite{Rovelli:2011eq}. 
$B$ is the momentum conjugate to the gravitational connection. It lives in the Lorentz algebra and generates local Lorentz transformations. Anytime a Lorentz frame (a time direction) is selected, $B$ can be decomposed into boost  and rotation parts.  For the boosts, we have
\ba
\vec K&:&
B^{oi}=\left[
\Big(\!\ast +\frac1\gamma  \Big) \Big(   e \wedge e \Big)
\right]^{oi}
\\ &=&
\frac12\epsilon^{oi}_{~\,jk} e^j \wedge e^k   +\frac1\gamma \ e^o \wedge e^i \ ,
\ea
where $i,j,k=1,2,3$ are space indices. For the rotation 
\ba
\vec L&:&
B^{ij}=\left[
\Big(\!\ast +\frac1\gamma  \Big) \Big(   e \wedge e \Big)
\right]^{ij}
\\ &=&
\frac12\epsilon^{ij}_{~\,ok} e^o\wedge e^k   +\frac1\gamma \ e^i \wedge e^k \ .
\ea
In particular, on a timelike surface with coordinates $z$ and $t$ and in the gauge where the tetrad is diagonal we have
\ba
K^i&=& \frac1\gamma \ e^o \wedge e^i \ ,
\ea
and
\vspace{-2em}
\ba
L^i&=&e^o\wedge e^i 
\ea
which shows that the generator of boosts and the generator of rotations are proportional. 

Consider now the area of the region $\cal R$ defined in the previous section, for an accelerated observer in a gravitational context. Assume the acceleration to be high so that $\cal R$ is small, and the tetrad can be considered constant on $\cal R$.   If we gauge fix the tetrad to a diagonal form, this is given by the integral over $\cal R$ of $e^o\wedge e^z$. Therefore in gravity we can write (in this gauge)
\be
  A=\int_{\cal R} \gamma K^z=\int_{\cal R}  L^z
  ~.
\ee
In the covariant quantum theory, these quantities are given by Lorentz generators on $\gamma$-simple unitary representations of $SL(2,{\mathbb{C}}
)$. But $L^z$ is a generator of a rotation subgroup of  $SL(2,{\mathbb{C}})$ and therefore has discrete spectrum in the quantum theory. Its eigenvalues are given by standard angular momentum theory as $m\hbar$ where $m$ is a half-integer and we are in units where the action is \eqref{action}, namely $8\pi G=1$ ($G$ is the Newton constant). Restoring physical units, we have a minimal nonvanishing value of the area 
\vspace{-1em}
\be
A_{min}=4\pi G \hbar ,
\ee
which, recalling \eqref{area}, gives a maximum physical value of the accelerations
\be
a_{max}=\sqrt{\frac1{8\pi G \hbar }}.
\ee
The existence of a maximum value of acceleration is of course something long expected in quantum gravity.  Here we have seen an indication on how it is realized in the loop theory.  Equivalently this gives a \emph{minimum} value for the horizon distance $\ell$
\be
\ell_{min}=\sqrt{{8\pi G \hbar }}.
\ee
which can be also viewed as an intrinsic uncertainly in the horizon position. Had we chosen a larger $\eta$, we would have obtained a weaker bound; a smaller $\eta$, on the other hand, does not make sense because it corresponds to a proper length $s=\eta/a$ larger than $\ell=1/\eta$, which is to say a path between $P$ and $P'$ shorter than the quantum fluctuations.

\section{Wedge amplitude}

Let us now study the actual covariant dynamics of the trajectory of the accelerated observer. This can be equally seen passively as a motion of an observer  in spacetime or actively as an evolution of spacetime seen by an observer.   To first order, the amplitude of this process is given by a single wedge amplitude \cite{Bianchi:2012ui}, where we can identify the region $\cal R$ with the wedge itself. The wedge amplitude is  \cite{Rovelli:2011eq} 
\be
W(g,g',h)=\sum_j (2j+1) ~ {{\rm Tr\,}}_{\!j}[Y^\dagger g'g^{-1}Yh]
\ee
where $g,g'\in SL(2,C)$, $h\in SU(2)$, $j$ is an half-integer labeling irreducible representations of $SU(2)$. See the reference cited for the rest of the notation and details. Here the product $g'g^{-1}$ can be taken to be precisely the boost between $P$ and $P'$, in the time gauge in both points. Therefore the amplitude reads 
\be
W(\eta,h)=\sum_j (2j+1)~ {{\rm Tr\,}}_{\!j} [Y^\dagger e^{i\eta K_z}Yh].
\ee
It is convenient to Fourier transform this from the group elements to the spin elements, which gives 
\be
W(\eta,j,m,m')=  \langle j,m|Y^\dagger e^{i\eta K_z}Y| j,m'\rangle.
\ee
The magnetic number refers to the orientation, which is not relevant here. Restricting to the $m=j$  coherent states, we have the amplitude 
\be
W(\eta,j)=  \langle j,j|Y^\dagger e^{i\eta K_z}Y| j,j\rangle.
\ee
This amplitude has been studied in \cite{Bianchi:2012ui}, where it is shown that its Fourier transform in $\eta$ peaks sharply on $\gamma j$ with a relative dispersion that decreases for large $j$.  Recalling that the spectrum of the energy can be read from the support of the Fourier transform in $t$ of the transition amplitudes, this can be taken as an indication of discreteness. A more detailed analysis of this amplitude will be given elsewhere. 

\section{Cosmology}

The resolution of classical singularities under the assumption of a maximal acceleration has been studied using canonical methods
for Rindler \cite{Caianiello:1989wm}, Schwarzschild \cite{Feoli:1999cn}, Reissner-Nordstrom \cite{Bozza:2000en}, Kerr-Newman \cite{Bozza:2001qm}  and Friedman-Lema\^itre \cite{Caianiello:1991gs} metrics.  Here we consider a simple homogeneous and isotropic cosmological model, with vanishing spatial curvature and pressure. The dynamics is governed by the Friedman equation
\be
\frac{\dot R^2}{R^2}= \frac{8\pi G}3 \rho 
\ee
where $R$ is the scale factor, $\dot R$ is its derivative with respect to proper time  and $\rho\sim R^{-3}$ is the matter energy density. Any comoving observer is accelerating with respect to his neighbors in this spacetime geometry.  Because of this, any observer has an horizon, at a distance
\be
\ell={R}/{\dot R}\ \ .
\ee
We have seen in the previous section that  the distance of the horizon is bounded by the minimal value $\ell_{min}$. Equivalently, the growing acceleration  approaching a classical singularity is bounded by the existence of a maximal acceleration $a\sim \sqrt{{\ddot R}/{R}}$. Still equivalently, this gives a maximal value of the energy density
\be
\rho_{\max}\sim \left.\frac3{8\pi G} \frac{\dot R^2}{R^2}\right|_{max}=\frac3{8\pi G} \ell_{min}^{-2}= \frac3{\hbar(8\pi  G)^{2}}.
\ee
So we recover a Planck-scale maximal energy density as in loop quantum cosmology  \cite{Ashtekar:2011ly}. The generic bound on acceleration implies that the resolution of cosmological singularities is general, supporting the results of \cite{Singh:2009mz,Singh:2010qa}.


\section{Conclusions}

Maximal acceleration appear to cure strong singularities (in the terminology of \cite{Tipler:1977kx,Krolak:1986ys}) such as  big bang, big crunch, black holes, as well as more exotic ones, such as big rip, in presence of violation of the strong energy condition. They should also be relevant for big-brake singularities, where the Universe suffers an infinite deceleration at finite size and zero velocity. Its relevance for other weak cosmological singularities, where only pressure diverges (sudden singularities), and other possible phenomenological consequences will be studied elsewhere. 

There are  models yielding a maximal acceleration by modifying the Lorentz transformations \cite{Magueijo:2001cr}.  We stress that the result here is compatible with conventional Lorentz invariance. Like for minimal length (and for the minimal value of an angular momentum component), also for maximal acceleration, under a Lorentz transformation an observer measures different probabilities for the same eigenvalues: symmetry transformations rotate states, not  a discrete spectrum \cite{Rovelli:2002vp,Rovelli:2010ed}. 

The evidence for maximal acceleration we have presented is partial: we hope it might open the door to more refined treatments. For this, a more detailed investigation of the mathematical properties of the wedge amplitude and its connection with physics is needed.  Maximal acceleration is to be expected in a quantum theory of gravity, it may provide the general link between the dynamics of the theory and generic general relativistic singularities and, perhaps with observations.  


\vskip5pt
 \noindent
{

{\bf Acknowledgments}  ~ F.V. acknowledges support from the Netherlands
Organisation for Scientific Research (NWO) Rubicon Fellowship Program.}


%

 \end{document}